\documentclass[twoside,11pt]{article}

\usepackage{doc_style_fixed}
\usepackage{hyperref,endfloat}

\heading{}{}{}{}{}{Foulkes, Thaweethai, and Reeder}

\ShortHeadings{Statistical Considerations in Long COVID Research}{Foulkes et al.}
\firstpageno{1}

\begin{document}

\title{Statistical Considerations in Long COVID Research}

\author{%
\name Andrea S. Foulkes \email afoulkes@mgh.harvard.edu \\
\addr Center for Biostatistics, Massachusetts General Hospital \\
Department of Medicine, Harvard Medical School \\
Department of Biostatistics, Harvard T.H. Chan School of Public Health \\
Boston, MA, USA 02114 
\AND \\
\name Tanayott Thaweethai \\
\addr Center for Biostatistics, Massachusetts General Hospital \\
Department of Medicine, Harvard Medical School  \\
Department of Biostatistics, Harvard T.H. Chan School of Public Health \\
Boston, MA, USA 02114
\AND \\ 
\name Harrison T. Reeder \\
\addr Center for Biostatistics, Massachusetts General Hospital \\
Department of Medicine, Harvard Medical School \\
Boston, MA, USA 02114
}

\maketitle

\begin{abstract}
Long COVID is a condition characterized by ongoing or relapsing symptoms attributable to SARS-CoV-2 infection that are present three or more months after infection. It represents a major clinical and public health concern as an estimated 5-10\% of individuals with a history of SARS-CoV-2 infection present with long term sequelae that range from mild to debilitating with profound impacts on quality of life. Clinical research studies of Long COVID have emerged rapidly over the past few years, and with them we are seeing several new data analytic challenges. In this manuscript, we highlight statistical challenges arising from the defining features of LC and associated study design strategies. This work is motivated by the Researching COVID to Enhance Recovery (RECOVER) Adult and Pediatric observational meta-cohort studies. 
\end{abstract}


\section{Introduction}
More than 700 million individuals worldwide have been infected with SARS-CoV-2 \citep{WHO}. Post-acute sequelae of SARS-CoV-2 infection (PASC) or Long COVID (LC), defined broadly as a condition characterized by ongoing, relapsing, or new symptoms due to infection present three or more months after infection, continues to be a major clinical and public health concern. Population-based prevalence estimates of long term, and often debilitating, symptoms after SARS-CoV-2 infection, are between 5-10\% with higher estimates associated with eras dominated by earlier variants \citep{Adjaye2023Long,Magnusson2023Prevalence}. Understanding the clinical course of LC, associated risk and resiliency factors, and the pathophysiology of persistence and recovery requires leveraging data from large observational cohort studies. Robustly designed clinical trials are needed to evaluate potential treatments for Long COVID. The statistical challenges inherent in these investigations are numerous due to the unique and complex data attributes and data generating processes in LC research. 

A fundamental challenge in LC research is the absence of a gold standard clinical definition or biomarker for LC. At this time, researchers must rely on patient self-diagnosis, the presence of a combination of self-reported symptoms, or evidence of clinician diagnosis in the medical record, which is sometimes (but not always) recorded as an ICD-10 diagnosis code for post-COVID condition \citep{Wang2026Post}. In the absence of clear diagnostic criteria, each of these indicators is inherently subjective and therefore prone to misclassification. This results in a familiar tradeoff between sensitivity and specificity, in which a highly sensitive definition may capture most individuals with LC but also incorrectly include individuals with other chronic conditions, while a highly specific definition may avoid false positives but also fail to capture many individuals with LC.  

Compounding the challenge of defining LC, an increasing body of evidence indicates that LC changes in presentation over time, with symptoms waxing and waning. In this sense, LC can be thought of as a latent state that manifests intermittently, making it challenging to identify individuals with LC. Moreover, LC presents differently for different people, with distinct symptom profiles that likely represent different pathological mechanisms. One approach that has been used to evaluate the mechanistic underpinnings of LC in this complex setting is to conduct clinical assessments at times when symptoms flare, by designing observational studies with repeated sampling opportunities that oversample individuals with active LC-related symptoms. Repeated sampling is needed to account for the waxing and waning of symptoms over time, and the consideration of LC-related symptoms allows for targeted assessments in individuals with specific LC sub-phenotypes.  Analysis of data derived from this design strategy, referred to as repeated auxiliary-variable dependent sampling, requires statistical innovation and novel methodological developments.   

In this article we first highlight current approaches to defining LC in the absence of a gold-standard, including one based on the concept of negative-unlabeled data that has been widely adopted (Section 2). We then describe defining attributes of LC that complicate clinical research studies of this condition, including its multiple and time-varying presentations (Section 3). Finally, with a lens focused on statistical challenges, we describe auxiliary-variable dependent sampling designs originally devised to accommodate ambiguity in LC and its multiple and time-varying presentations (Section 4).

\section{Defining a Syndrome Without a Gold Standard}
Defining LC in the absence of gold standard diagnostic criteria is a statistical conundrum. Supervised learning approaches, such as generalized linear modeling, penalized regression, and ensemble learning (random forests, boosting), allow for identifying symptoms associated with LC or an algorithm for determining LC status based on a combination of features; however, these methods require an existing LC definition to label each data point for training. Unsupervised learning, such as hierarchical or k-means clustering, can be used to identify groups of individuals with distinct patterns of symptoms or other clinical characteristics, but cannot determine which of these clusters correspond to LC. Moreover, in the absence of a gold standard, existing validation techniques \citep{Steyerberg2001Internal} for classification rules are not easily applicable. 

Conceptually, we can consider LC status as negative-unlabeled data, in recognition that individuals without a history of SARS-CoV-2 infection are “negative” for LC (by definition), while those with a history of infection are “unlabeled” (i.e., positive or negative) for LC. A related phenomenon with an emerging literature is positive-unlabeled data \citep{Bekker2020Learning,Liu2003Building,Liu2025Positive,Van2020survey} which is ubiquitous in electronic health records (EHR) data settings, occurring when individuals may have “positive” documentation of a medical condition, but lack of documentation does not guarantee lack of the condition. However, while it is conceptually similar to positive-unlabeled data, negative-unlabeled data has received far less attention.

\begin{figure}
\caption{Use of infection status as a pseudo-label for LC in the absence of a gold sstandard} \label{fig:pseudolabel}
\smallskip

\includegraphics[scale=0.5]{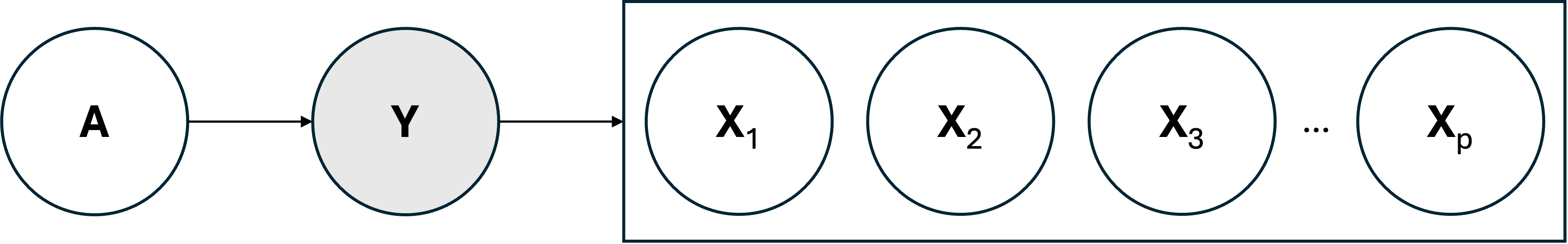}
\bigskip

$A =$ infection history; $Y =$ Latent Long COVID Status; $(X_1, X_2, X_3,\dots, X_p)=$ reported symptoms. $Y$ is unobservable (indicated by grey shading) in the absence of a definition or biomarker for LC. 
Variables $X_1, X_2, X_3, \dots, X_p$ are not precursors of $A$, and the effect of $A$ on  $X_1, X_2, X_3, \dots, X_p$ is through the potential emergence of $Y$. 
\end{figure}

To develop a classifier based on current symptoms or other clinical characteristics to distinguish people with and without LC, one strategy is to train the model using a binary “pseudo-outcome” indicator for history of SARS-CoV-2 infection. This approach results in a decision rule that differentiates individuals who have never been infected from those who have a history of infection (Figure~\ref{fig:pseudolabel}). An example of such an approach was the derivation and application of the LC Research Index (LCRI). The LCRI was derived by first applying Lasso-penalized logistic regression using infection history as the dependent variable and self-reported symptoms as the independent variables \citep{Thaweethai2023Development,Geng20242024,Gross2025Characterizing,Gross2024Researching,Reeder2025Penalized}. From this model, nonzero symptom coefficients were converted into a numeric index corresponding to an individual’s current symptoms, which is referred to as the LCRI. In turn the LCRI was thresholded to classify individuals with LC. 

For example, the adult LCRI (originally published in \citet{Thaweethai2023Development} and updated in \citet{Geng20242024}) involves evaluating an individual for the presence of 11 symptoms, where each symptom was assigned a score from 1 to 7. The LCRI is calculated by adding up the scores for the symptoms that are present and checking against a total score threshold of 11 for LC. This decision rule was identified to control false positives within individuals without a history of infection. While inherently subject to measurement error, this approach is the only strategy to our knowledge that rigorously defines LC such that misclassification of chronic conditions arising for reasons other than SARS-CoV-2 infection is minimized, thus ensuring the specificity of the definition. Additional LCRI have been derived in pediatric populations in a similar fashion \citep{Gross2024Researching,Gross2025Characterizing}.

The LCRI threshold approach, however, can be subject to false negatives \citep{Azola2025Putting}, particularly individuals who have few symptoms that add to the index, but severe symptom burden nonetheless. One analysis described to reduce false negatives is to focus only on individuals who meet the LCRI threshold for LC and  individuals with an LCRI of 0 (i.e., report no symptoms that contribute to the LCRI) \citep{Erlandson2024Differentiation}. Such an analysis treats individuals with an LCRI greater than 0 but less than the LCRI threshold for LC in a separate “unclassified” category. A greater share of false negatives are expected in the unclassified category compared to those with an LCRI of 0. A comparison to individuals with LCRI equal to 0 is preferred to a comparison to those with no symptoms at all (inclusive of symptoms not contributing to the LCRI), as individuals with no symptoms at all tend to be healthier than the general population. Indeed, even SARS-CoV-2 uninfected individuals tend to report symptoms, and individuals with no symptoms at all do not constitute a good comparator group for most analyses.  

An important complementary strategy that maximizes sensitivity is the National Academy of Science, Engineering, and Medicine (NASEM) characterization of LC \citep{Ely2024Long} which is designed to be broadly inclusive. According to the NASEM findings, LC is an “infection-associated chronic condition (IACC) that occurs after SARS-CoV-2 infection and is present for at least 3 months as a continuous, relapsing and remitting, or progressive disease state that affects one or more organ systems”. Under this definition, LC encompasses the presence of any symptoms matching this description that are new or worsened following SARS-CoV-2 infection. However, in a research setting, and particularly in a large observational study that relies on self-reported symptoms and health status, clinical evaluation to ascertain a link between new/worsened symptoms and SARS-CoV-2 infection is not generally feasible. 

In the absence of a clinical evaluation, defining LC based on the presence of any self-reported symptoms will likely minimize false negatives at the expense of potentially concluding individuals have LC when in fact they have another chronic condition. This type of low-specificity definition can lead to a reduction in statistical power and an attenuation of effect sizes between LC and biomarkers (or other outcomes) due to measurement error \citep{Thomas2025EXPOSURE} and is therefore generally not recommended for research purposes. Indeed, in the section of the NASEM report titled “How Can Researchers Apply the Definition?” it is acknowledged that researchers may need to apply additional criteria beyond their stated definition to ensure sufficient specificity to meet study objectives \citep{Ely2024Long,National2024Long}.

Notably, in LC research studies, LC may be considered the exposure, e.g., in evaluating the impact of LC on a clinical assessment, or the outcome, e.g., in determining risk factors for LC or time to recovery from LC. Drawing on the definitions described above, LC can be measured as a quantitative trait (e.g., the LCRI), a binary indictor for the presence of a condition (e.g., LCRI $\ge 11$ vs. LCRI $<11$), a categorical variable (e.g., LCRI $\ge 11$ vs. $0<$ LCRI $<11$ vs. LCRI $=0$), or a time-to-event outcome (e.g., time to recovery from LC). Moreover, additional information on specific symptoms or LC sub-phenotypes can be integrated into any exposure or outcome definition as described in the next section.

An alternative approach that is intuitively appealing due to its ease of implementation uses a count of all reported symptoms as the outcome. Importantly, a symptom count is sensitive to the number of symptoms in the data collection instrument, how these symptoms distribute across organ systems, and the within organ system correlations. For example, the count of symptoms for someone with a cardio-pulmonary manifestation of LC may differ from the count of symptoms for someone with a neurological manifestation if the number of recorded symptoms or the correlation of these symptoms differs between the two domains. Methods for determining the effective number of tests have been described, for example in the evaluation of gene-level associations where each gene is comprised of multiple correlated SNPs \citep{Cheverud2001simple,Li2005Adjusting, Nyholt2004Simple}, and could be extended to this setting. In our prior work we demonstrate through simulation studies that a naïve count of symptoms performs considerably worse than the LCRI with respect to discriminative performance \citep{Reeder2025Penalized}.

\section{Multiple and Time-Varying LC Conditions}
A further challenge in LC research is that it manifests differently across individuals and over time. Several reports suggests that there are distinct sub-phenotypes of LC associated with multi-organ dysfunction including pathologies in the blood vessels, brain, heart, gastrointestinal tract, immune system, kidneys, liver, lungs, pancreas, reproductive system, and spleen \cite{Davis2023Long}, and characterized by different symptom profiles \citep{Gross2025Characterizing,Gross2024Researching,Geng20242024,Thaweethai2023Development,Wang2026Post}. While acknowledging the challenges described previously to “defining” LC via unsupervised learning, there have been several attempts using Bayesian clustering and k-means clustering methods to characterize different patterns in prevalent symptomatology following SARS-CoV-2 infection, that may represent various LC sub-phenotypes \citep{Wang2025Refinement}.

Alternatively, LC sub-phenotypes can be defined clinically based on specific symptom patterns. For example, a cardiopulmonary phenotype may be characterized by shortness of breath, postural tachycardia, or chest tightness or pressure while a neurological phenotype may be characterized by brain fog or difficulty communicating \citep{Wang2025Refinement}. Sub-phenotype groupings based on specific symptoms may be applied in research settings in combination with thresholding based on the LCRI. Use of symptom-defined sub-phenotypes alone is subject to many of the same drawbacks described in Section 2 for using the NASEM definition or a count of symptoms. That is, an individual with multiple cardiopulmonary symptoms may not necessarily have LC and could be a false positive if classified as LC. 

Notably, assignment of individuals symptom-based sub-phenotypes based on algorithmic clustering methods results in individuals only belonging to a single sub-phenotype group, whereas assignment to LC sub-phenotypes based on clinically defined symptom patterns may result in overlapping sub-phenotype groupings. For example, an individual who has a cardiopulmonary phenotype, e.g., shortness of breath, may or may not also present with a neurological phenotype, e.g., brain fog. In either case, approaching comparative analyses across sub-phenotype requires appropriately defining comparator groups in the statistical analysis, with careful consideration of the scientific questions at hand. Additionally, if the sub-phenotype groups overlap, mixed membership models \citep{2014Handbook} or methods for partially overlapping samples \citep{Derrick2017Test} are needed to account for individuals belonging to more than one group \citep{Rao2024Postacute}.    

In addition to presenting in many potential forms or sub-phenotypes, LC can also manifest differently in the same individual over time. That is, an individual may initially present with LC characterized primarily by cardiopulmonary symptoms, and later present with primarily neurologically or gastrointestinal symptoms. This variation may be due to a reinfection, perhaps with a new variant, or could represent the natural progression of the disease, especially in pediatric populations as they progress through different phases of childhood and adolescence. In addition, symptoms tend to ebb and flow over time \citep{Thaweethai2025Long}, and such fluctuations may or may not represent a change in the underlying disease state. That is, an individual may have a symptom associated with LC, e.g., post-exertional malaise (PEM), that comes and goes periodically, which may represent recovery, the natural history of LC, or a temporarily resolution of symptoms due to an external cause. In other words, the absence of PEM at a given study visit (and an associated dip in the LCRI) does not necessarily indicate a resolution of LC as it may return at a later visit if the underlying condition is still present. In light of these phenomena, possible research definitions of ``recovery'' include a fold-reduction in the LCRI, e.g., a 50\% or 75\% decrease, and/or one or more consecutive indications of no LC, e.g., LCRI less than the threshold, over a specified time, e.g., a 6-month period, but a scientific and clinical consensus on what constitutes LC resolution is needed. 

\begin{figure}
\caption{RECOVER-Adult Cohort design.} \label{fig:adultdesign}
\bigskip

\centerline{
\includegraphics[scale=0.5]{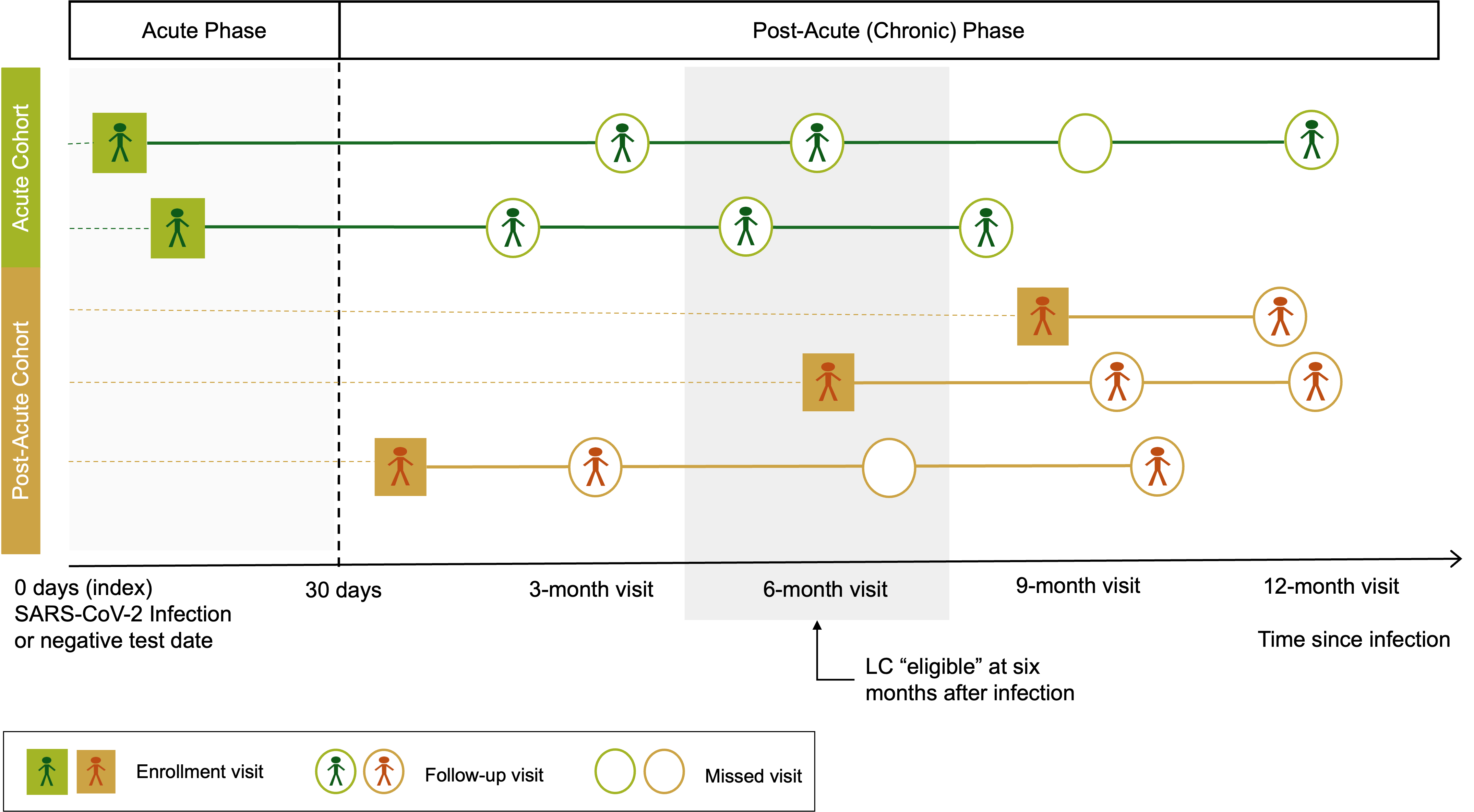}
}
\bigskip

In the RECOVER-Adult design, participants belong to one of two sub-cohorts – the Acute Cohort or the Post-Acute Cohort – depending on whether they enrolled within 30 days of index date. As this design allows for enrollment at varying times from SARS-CoV-2 infection, some participants enroll more than 6-months after first infection, the first timepoint at which they were eligible to be classified as having LC. In these cases, LC status is subject to structural intermittent missingness.          
\end{figure}

Further complicating studies of LC is the fact that it can be unobservable over an interval of time, e.g., after a recent reinfection when LC can be masked by acute phase symptoms. Thus, new onset, continued presence, and resolution of LC are subject to structural intermittent missingness during the acute phase of a reinfection. Finally, the rapid emergence of SARS-CoV-2 led many prospective cohort studies to enroll individuals who already had infection and emergent LC symptoms, resulting in LC onset that is left censored by design. For example, in the Researching COVID to Enhance Recovery (RECOVER) Adult cohort study, a substantial proportion of individuals were enrolled well after the acute phase of infection, as depicted in Figure~\ref{fig:adultdesign}. As a result, for these individuals we do not know if LC presented as a continuation of symptoms experienced during the acute phase of infection or if it began several months to a couple of years after initial infection. Furthermore, for individuals without LC at the time of study entry, we do not know if they previously experienced the condition but also recovered prior to enrollment. 

This final challenge is further complicated by the potential association between left censorship and viral variant. In RECOVER-Adult, for example, the time between first infection and enrollment was positively associated with likelihood of first infection being an early (pre-Omicron) variant. In fact, almost all study participants enrolled during the acute phase of infection were first infected after December 1, 2021, when the Omicon variant became dominant across the United States. As a result, the population represented by studies at early time points (in relation to first infection) may be fundamentally different than the population represented in studies of later time points.               

\section{Auxiliary-Variable Dependent Sampling}

To address ambiguity in LC and its multiple and time-varying presentations, LC research studies have relied on auxiliary-variable dependent sampling techniques. Such designs exemplify a more general methodology referred to as two-phase sampling, in which inexpensive variables (collected in phase one) are used to inform the measurement of more expensive variables (collected in phase two). Methodologies for addressing the selection bias introduced by such designs have been widely studied \citep{Breslow2003Large,Breslow1997Maximum,Hejazi2021Efficient,Kennedy2020Efficient,Robins1994Estimation,Rose2011targeted,Rotnitzky1995Semiparametric,Wang2009Causal}. 

\begin{figure}
\caption{Auxiliary-variable dependent sampling at a fixed study visit.} \label{fig:auxsamp1}

\bigskip
\centerline{
\includegraphics[scale=0.8]{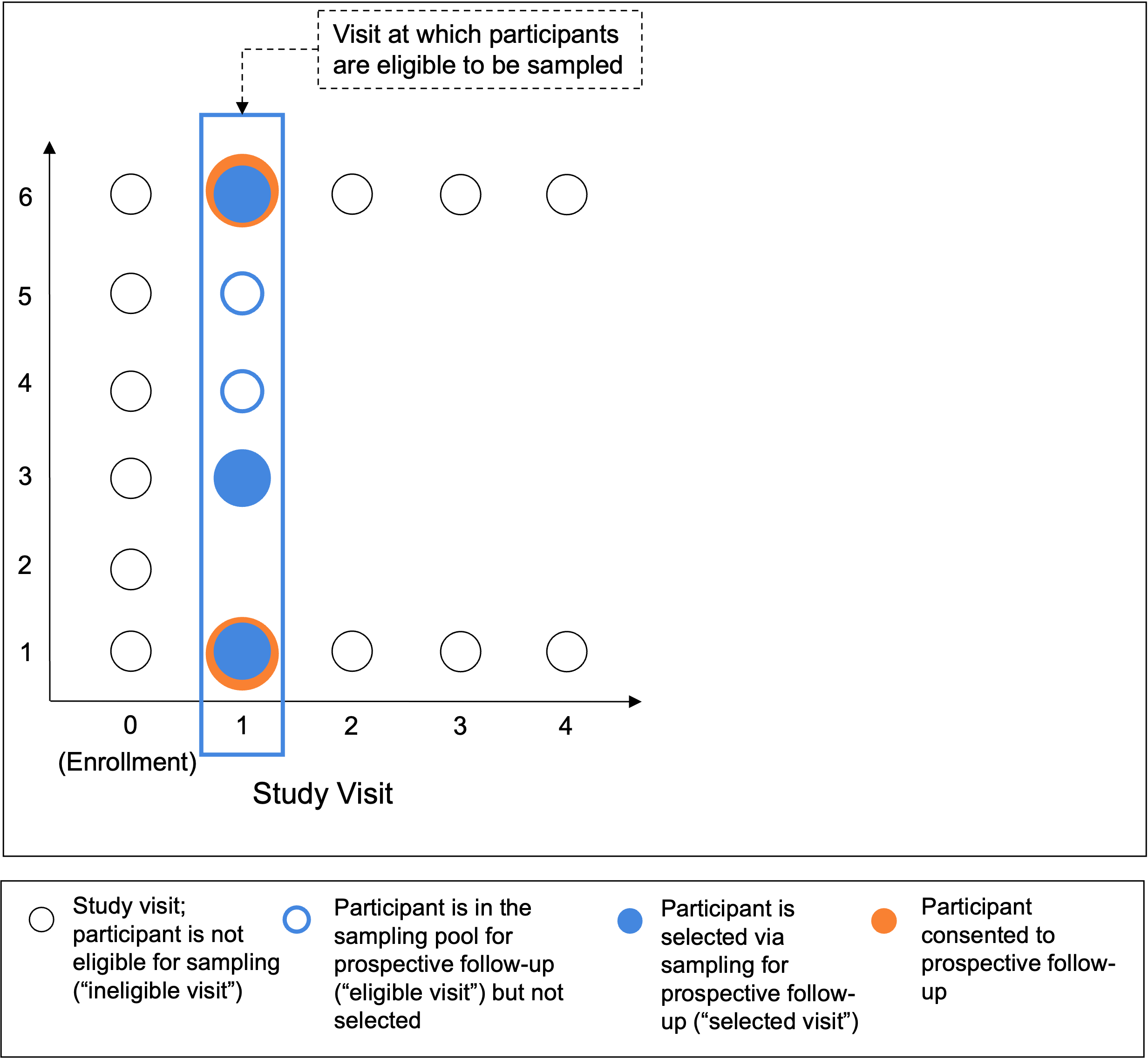}
}

\bigskip
In this example, participants are selected via sampling at a fixed study visit and then continue for prospective follow-up, like in the RECOVER-Pediatrics cohort.
\end{figure}

\begin{figure}
\caption{Repeated auxiliary-variable dependent sampling.} \label{fig:auxsamp2}

\bigskip
\centerline{
\includegraphics[scale=0.8]{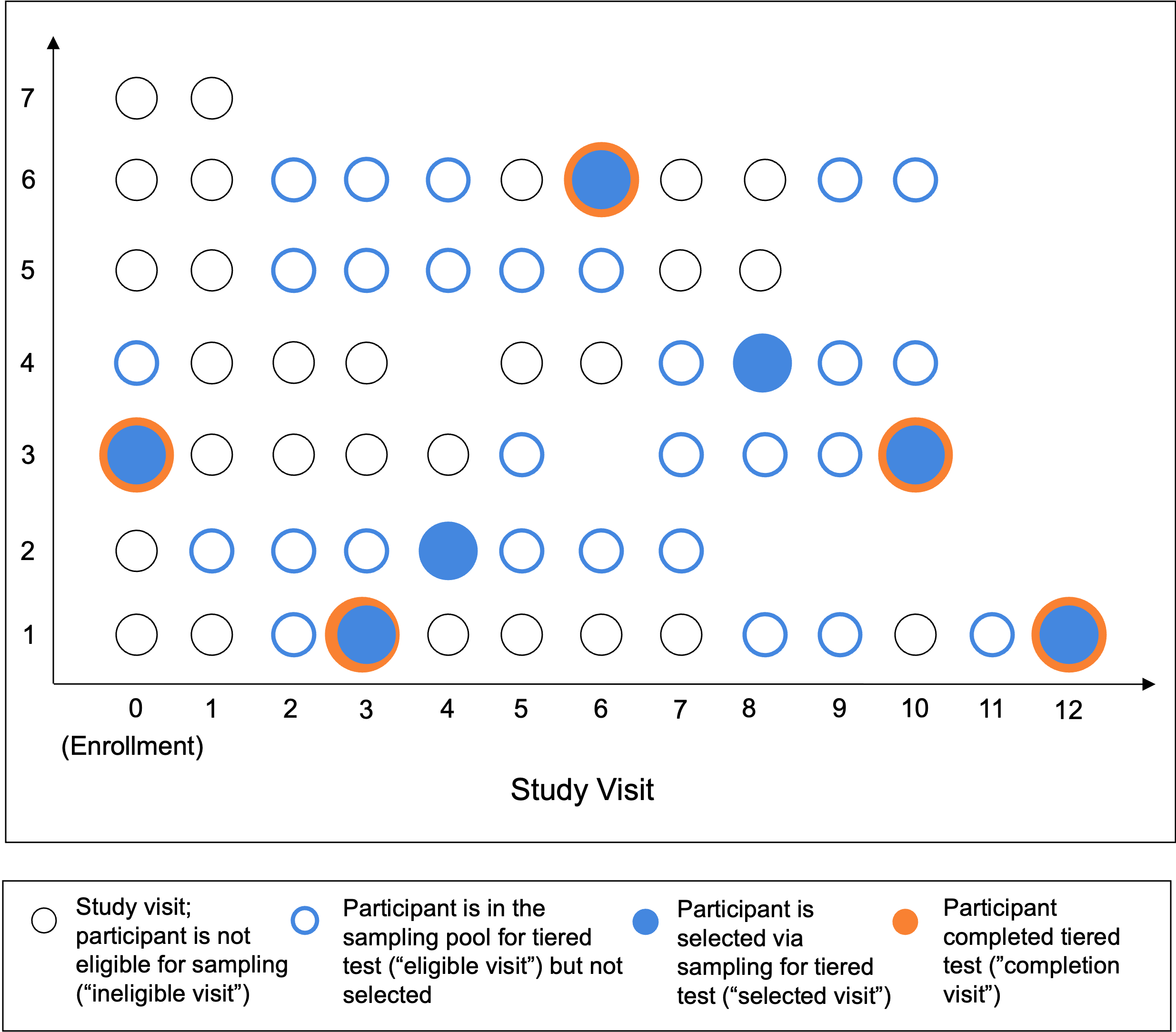}
}

\bigskip
In this complex design, sampling occurs repeatedly and at variable study visits depending on eligibility requirements, like in the RECOVER-Adult cohort.
\end{figure}

Visual depictions of the design of LC research studies involving auxiliary-variable dependent sampling are provided in Figures~\ref{fig:auxsamp1} and~\ref{fig:auxsamp2}. We refer to assessments performed only in a subset of individuals as “tiered” tests, to distinguish from those performed in the whole cohort. In the simplest design setting (Figure~\ref{fig:auxsamp1}), study participants reach a pre-defined study visit at which time they may be selected to receive an additional assessment based on a randomization schema that is dependent on one or more variables. For example, in the RECOVER-Pediatrics Cohort study, an initial remote visit was performed for all enrollees after which only a subset of participants oversampled for those with prevalent LC symptoms underwent longitudinal follow-up \citep{Gross2024Researching}.

More complex designs (Figure~\ref{fig:auxsamp2}) incorporate sequential opportunities to sample participants for additional assessment, based on auxiliary variables that can vary from visit to visit. For example, in the RECOVER Adult cohort study, at each visit participants could be selected for additional assessments, with selection probabilities based on the values of auxiliary variables at that visit \citep{Horwitz2023Researching}. In this study, test administration is further limited to participants meeting eligibility requirements, involving, for example, time since most recent SARS-CoV-2 infection and time since a prior study visit when the test was completed. 

Such a design exemplifies general two-phase strategies in which repeated sampling depends on longitudinal eligibility requirements and auxiliary variables, as well as the result of sampling at prior visits. In total, RECOVER-Adult included administration of over 50,000 tests over 31 different types of tests based on a repeated auxiliary-variable sampling schema. The specific scheme used in RECOVER-Adult set the selection probability to 1 for individuals with specific ‘trigger’ variables present at a particular visit, and a low rate of 0.056 otherwise. A few specific examples of RECOVER-Adult test triggers corresponding to select tiered assessments are provided in Table~\ref{tab:triggers}.

\begin{table}
\caption{Sample triggers for tiered tests in RECOVER-Adult} \label{tab:triggers}
\centerline{
\includegraphics[scale=0.75]{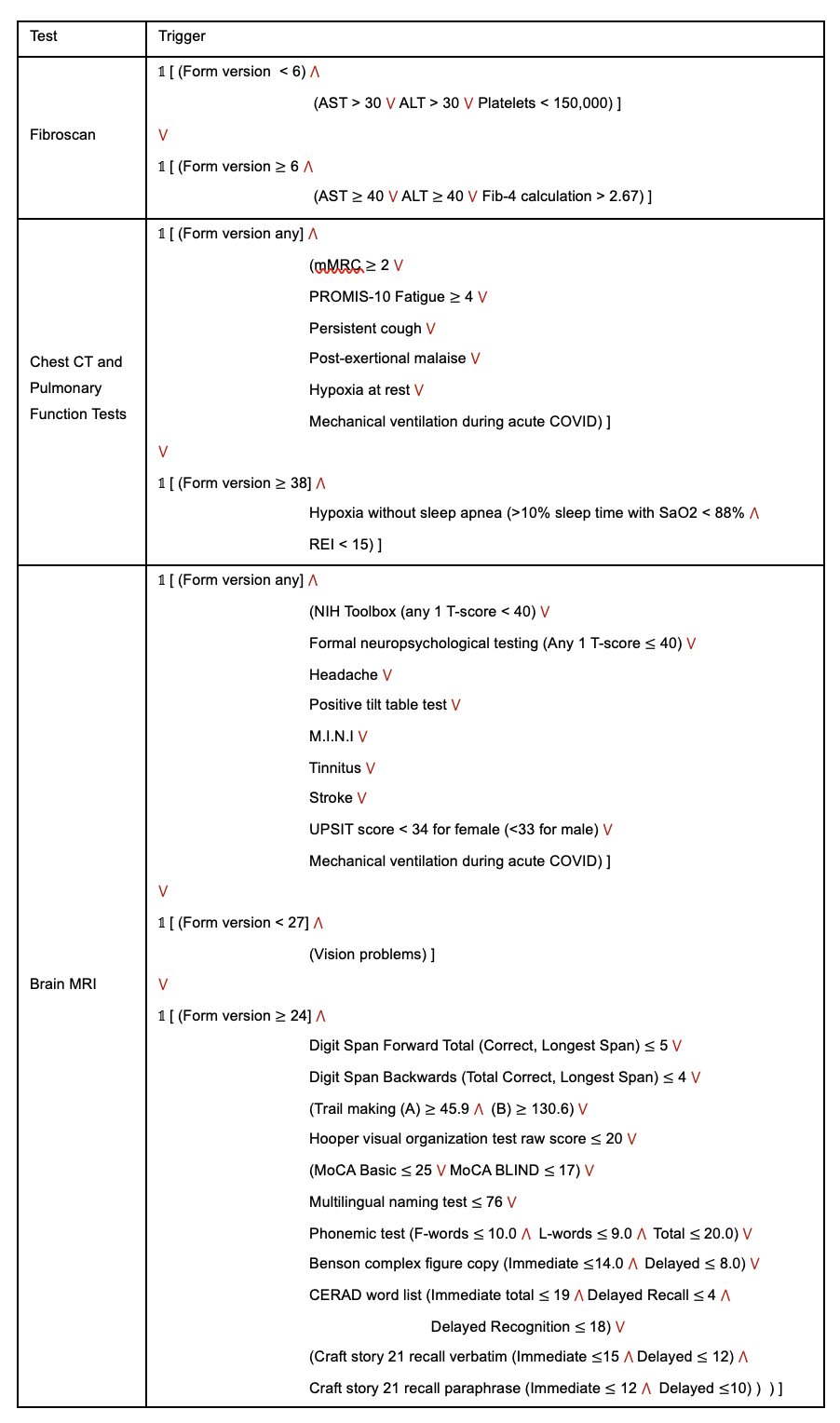}
}

\footnotesize{$\vee$ indicates ``or" \\ $\wedge$ indicates ``and"}
\end{table}

Auxiliary variables used in defining sampling criteria tend to be leveraged precisely because they are expected to be associated with the results of a tiered test outcome. Without accounting for the sampling design, analyses of association between an exposure and the tiered test outcomes are subject to selection bias due to differences between the sampled group and the cohort overall. For example, if the exposure of interest truly is associated with worse health outcomes, but the auxiliary variable is also a marker of poor health, then the sampling mechanism may select most of the exposed individuals and only the sickest unexposed individuals. The result would be an observed bias toward the null of no association because both the exposed and non-exposed participants would be differentially oversampled for participants with characteristics associated with poor tiered test results. Moreover, even among people who are sampled, missingness of the outcome in the phase-two sample can further induce selection bias. Participants who are sampled to receive a tiered test may not always complete it, resulting in informative missingness that additionally needs to be accounted for in the analysis to avoid selection bias.

Beyond the complex sampling scheme used to collect tiered test results, analytic approaches also must address the typical challenges inherent to observational prospective cohorts, such as confounding and dropout. For investigations treating LC as a time-varying exposure, meaningful imbalances across exposure groups can lead to confounding. In general, in observational studies, exposure is not randomized, and therefore inferences could be compromised by exposure-outcome confounding and ignoring imbalance across exposure groups can lead to spurious associations. The presence of time-varying confounding renders this setting distinct from those in which the exposure is randomized, such as clinical trials for which analytic methods for auxiliary variable dependent sampling have been previously explored \citep{Gilbert2014Optimal}.

In addition, loss to follow-up prior to the sampling event presents a final source of potential selection bias. Participants can be lost to follow-up between enrollment and the first visit they would have been eligible to be sampled for a tiered test. Ignoring information on such participants in the analysis can lead to biased or non-generalizable findings if there are systemic differences between participants lost to follow-up in this period, compared to participants who reached the first sampling event. For example, if these individuals ended participation due to death or severe illness, then any conclusions about the association between exposure and the tiered test outcome would be limited to relatively healthy individuals represented by participants who reached the first sampling event.

Finally, it is important to note that the temporal definition of exposure matters. Research questions involving tiered tests generally focus on whether the results of the test vary by levels of an exposure. Clarity on the temporal relationship between this exposure, the sampling event, and test administration is essential for appropriate application of analytic tools. 

\section{Discussion}
A summary of the data analytic challenges in LC clinical research highlighted above is provided in Table~\ref{tab:summary}. Recognizing the critical need to address these statistical challenges in RECOVER as well as the enormity of unanswered scientific and clinical questions, the National Institutes of Health allocated \$9.8M over three years (2024-2027) to establish a national Network of Biostatisticians for RECOVER (NBR). Led by the RECOVER Data Resource Core (DRC) at Mass General Brigham, the principal goal was to establish eight independent teams of biostatisticians within NBR to lead statistical analyses that leverage the full array of RECOVER data. The specific objectives were for NBR Biostatisticians to: 1) learn the complexities of RECOVER observational cohort data, including study design and data considerations through direct, hands-on tutorials and working groups; 2) participate in forums for discussing, developing, and disseminating best practices for data analysis, including developing standard operation procedures for principal definitions and distributing open-source code and training materials; and 3) collaborate with RECOVER DRC and the RECOVER Adult, Pregnancy, Pediatrics, and Pathobiology Coordinating Committees on manuscript writing committees to effectively achieve identified scientific priorities. Members of NBR were additionally expected to engage in developing novel statistical methods to address the unique challenges of RECOVER data with the foremost goal of advancing the scientific priorities of the RECOVER consortium.

\begin{table}
\caption{Summary of important statistical considerations in LC clinical research} \label{tab:summary}
\centering
\includegraphics[scale=0.6]{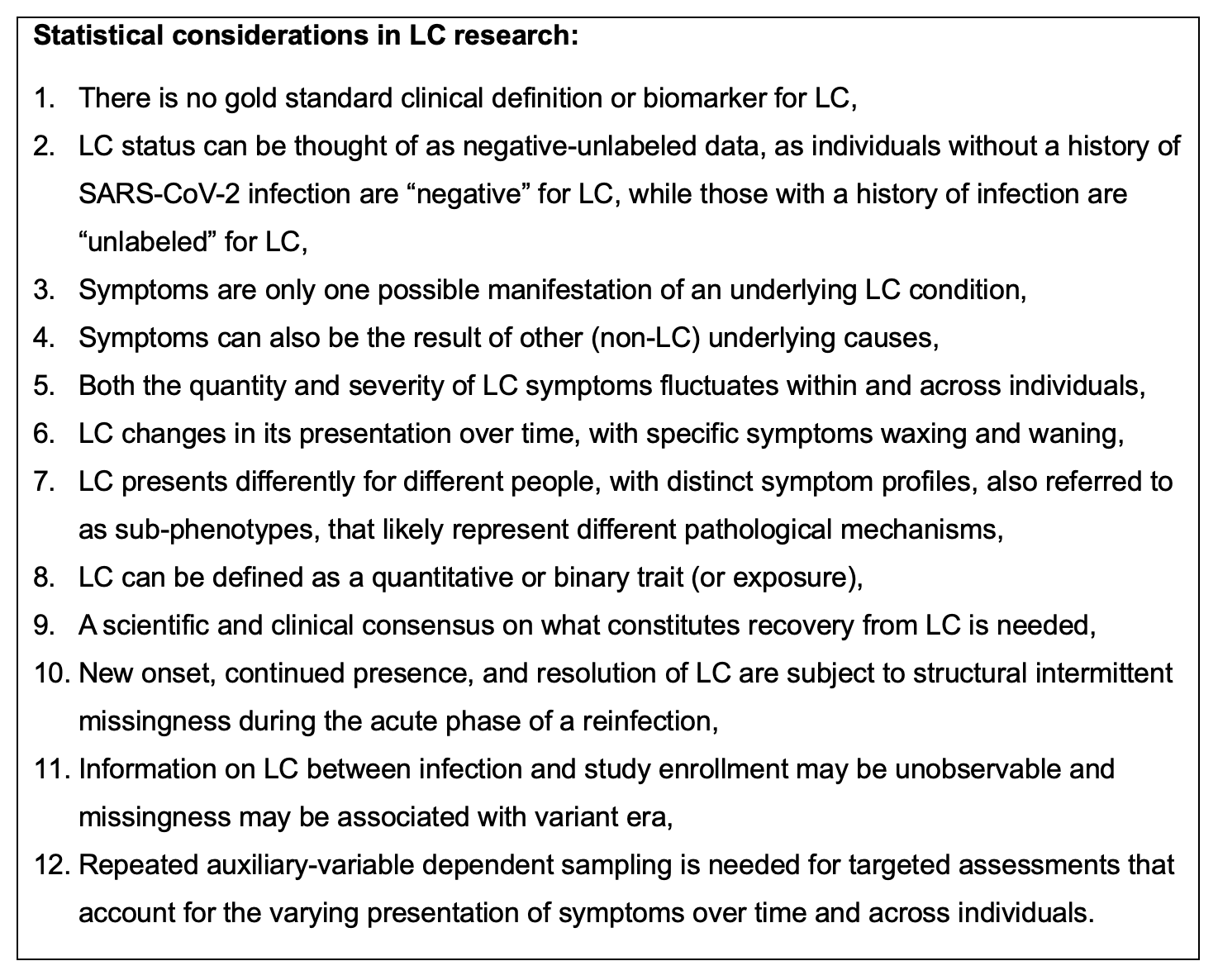}
\end{table}

While the concepts outlined in this manuscript are part and parcel of LC research, many of them apply to other research settings. More broadly, prioritizing testing based on data collected on auxiliary variables is a potentially cost-efficient design strategy for clinical trials and epidemiologic studies, including case-cohort and case-control designs \citep{Breslow1996Statistics,Breslow2009Using,Prentice1986case}, electronic health records (EHR)-based studies \citep{Amorim2021Two,Levis2024Double,Zhang2024Patient} , and immune correlate analyses of vaccine efficacy trials to measure candidate surrogate endpoints \citep{Follmann2006Augmented, Fong2015Calibration,Fu2017Joint,Gilbert2024Four}. Resource-efficient cohort study designs an optimal subset of participants may be selected for additional follow-up using information on case status \citep{Gross2024Researching}. In two-phase vaccine trials, an inexpensive auxiliary variable may be measured at a pre-defined study time point and an optimal subset may be selected for measuring a more expensive outcome \citep{Gilbert2014Optimal}. 

Appropriately leveraging the data generated from clinical studies of LC will undoubtedly yield immense and novel scientific discoveries, ultimately leading to significant strides in clinical care and prevention efforts. However, LC studies and associated data capture, present several analytic challenges and require rigorous and robust biostatistical analyses. Readers are encouraged to learn more about NBR and methods for analysis of RECOVER data at \href{https://www.recover-nbr.org/}{https://www.recover-nbr.org/}.        

\section*{Data Availability}
NHLBI has undertaken a significant effort to release harmonized data from all RECOVER observational cohort studies to the public via the \href{https://biodatacatalyst.nhlbi.nih.gov/recover}{BioData Catalyst platform}. Data are available for researchers whose proposed use of the data has been approved for a specified scientific purpose and after approval of a proposal and with a signed data access agreement.

\acks{Support for this research was provided by NIH OTA OT2HL161841 and R01 HL162373.}

\newpage

\vskip 0.2in
\bibliography{ref}

\end{document}